\documentstyle[prd,aps,epsfig]{revtex}

\author{S. Hegyi}

\title{H-function extension of the NBD
in the light of experimental data}

\address{KFKI Research Institute for Particle and Nuclear Physics \\ of the
	Hungarian Academy of Sciences, \\ H-1525 Budapest 114, P.O. Box 49.
	Hungary}

\date{\today}

\begin{document}

\pagestyle{plain}
\pagenumbering{arabic}

\maketitle

\begin{abstract}
  The recently introduced H-function extension of the Negative
  Binomial Distribution is investigated. The analytic form of $P_n$ is
  rederived by means of the Mellin transform.
  Applications of the HNBD are provided using experimental
  data for $P_n$ in $e^+e^-$, $e^+p$, inelastic $pp$ and
  non-diffractive $p\bar p$ reactions.
\end{abstract}


%\narrowtext

In multiparticle phenomenology
we celebrate the 25th anniversary year of the invention of two ideas
which turned out to be highly influential: the Koba-Nielsen-Olesen (KNO)
scaling of the multiplicity distributions $P_n$~[1]
and their Negative Binomial
Distribution (NBD) type parametrization~[2].
Both subjects received a great deal of attention  during the past 25 years,
see refs.~[3-5] for review articles.

A popular way of introducing the NBD as a model of $P_n$
is the Poisson transform defined by
\begin{equation}
	P_n=\int_0^\infty\psi(z)\,\frac{z^n}{n!}\,e^{-z}\,dz.
\end{equation}
For the NBD the continuous probability density $\psi(z)$ is the
gamma distribution providing the asymptotic KNO scaling form
of $P_n$~[3]. At high energies the agreement between the NBD and
observations is often not quite satisfactory therefore
it is of interest to look for possible generalizations.
We shall consider here an extension of the NBD
that can be obtained via the scaling transformation
$z\to z^{1/\mu}$ of the gamma variate
with scaling exponent $\mu>0$. The modified $\psi(z)$ can be written
in the form
\begin{equation}
        \psi(z)=\frac{\mu}{\Gamma(k)}\,\lambda^{\mu k}
        z^{\mu k-1}\exp\left(-[\lambda z]^\mu\right)
\end{equation}
which is the generalized gamma distribution~[6]
with shape parameter $k>0$ and scale parameter $\lambda>0$.
Its moments are given by
\begin{equation}
	\langle z^q\rangle=\int_0^\infty z^q\psi(z)\,dz=
	  \frac{\Gamma(k+q/\mu)}{\Gamma(k)}\frac{1}{\lambda^q}.
\end{equation}
For $\mu=1$ the ordinary gamma distribution is recovered and the Poisson
transform of Eq.~(2) yields the NBD,
\begin{equation}
	P_n=\frac{1}{n!}\,\frac{\Gamma(k+n)}{\Gamma(k)}\,
	\left(\frac{\langle n\rangle}{k}\right)^n
	\left(1+\frac{\langle n\rangle}{k}\right)^{-k-n}
\end{equation}
where $\langle n\rangle$ is the average multiplicity.
For $\mu\neq1$ the basic properties of the
Poisson transformed generalized gamma density  have been
determined in two previous Letters~[7,8]. Since the analytic form of
$P_n$ can be expressed in terms of the H-function of Fox~[9]
(see the Appendix for a summary) we will call the
distribution as HNBD for short. The main goal of the present paper is
the analysis of various multiplicity data
which are known to be in disagreement with the $\mu=1$ special case.

In ref.~[7] the analytic form of the HNBD was obtained utilizing the
Laplace transform of the Fox function given by the equation pair~(A.6-7).
This is a straightforward method since $n!P_n$ in Eq.~(1) is the Laplace
transform of the product $z^n\psi(z)$ which can be expressed in terms of
${\sf H}(x)$ for the generalized gamma distribution.
To illustrate the simplicity of manipulating H-functions, first we
provide a slightly different approach to the analytic form of the
newly developed probability law.
The Laplace and Mellin transforms of a function $f(x)$, $x\in(0,\infty)$,
are defined by
\begin{equation}
	{\cal L}(f(x),r)=\int_0^\infty e^{-rx}f(x)\,dx
\end{equation}
and by
\begin{equation}
	{\cal M}(f(x),s)=\int_0^\infty x^{s-1}f(x)\,dx
\end{equation}
respectively. They are related to each other through
\begin{equation}
	{\cal M}\{{\cal L}\,(f(x),r),s\}=\Gamma(s)\,{\cal M}\{f(x),1-s\}
\end{equation}
which can be justified by Fubini's theorem~[10]. From
the moments of $\psi(z)$ given by Eq.~(3) one obtains
\begin{equation}
	{\cal M}\{z^n\psi(z),s\}=\frac{\Gamma(k+(s+n-1)/\mu)}{\Gamma(k)}\,
	\frac{1}{\lambda^{s+n-1}}
\end{equation}
for the Mellin transform of $z^n\psi(z)$ and the use of Eq.~(7) yields
\begin{equation}
	{\cal M}\{{\cal L}\,(z^n\psi(z),r),s\}=
	\frac{1}{\lambda^n\Gamma(k)}\,
	\Gamma(s)\,\Gamma(k+n/\mu-s/\mu)\,\lambda^s.
\end{equation}
Comparison with the Mellin transform (A.8) of the Fox function shows
that ${\cal L}\,(z^n\psi(z),r=1)$, providing $n!P_n$, is expressible
in terms of ${\sf H}(x)$ with the following parameters:
\begin{equation}
   \int_0^\infty e^{-z}z^n\psi(z)\,dz=
   \frac{1}{\lambda^n\Gamma(k)}\;{\sf H}^{1,1}_{1,1}
   \left[
      \,\frac{1}{\lambda}\left|
      \begin{array}{c}
         (1-k-n/\mu,\; 1/\mu)         \\
         (0,1)
      \end{array}
   \right]\right.
   \quad\mbox{for }\mu>1
\end{equation}
where the restriction on $\mu$ follows from the existence condition
$\kappa>0$ of ${\sf H}(x)$ discussed in the Appendix.
Thus, using identity (A.4), one arrives at
\begin{equation}
   P_n=
   \frac{1}{n!\,\Gamma(k)}\;{\sf H}^{1,1}_{1,1}
   \left[
      \,\frac{1}{\lambda}\left|
      \begin{array}{c}
         (1-k,\; 1/\mu)         \\
         (n,1)
      \end{array}
   \right]\right.
   \quad\mbox{for }\mu>1
\end{equation}
and with the help of identity (A.3) we can write the analytic form of
$P_n$ in terms of a legal H-function also for $\mu<1$:
\begin{equation}
   P_n=
   \frac{1}{n!\,\Gamma(k)}\;{\sf H}^{1,1}_{1,1}
   \left[
      \,\lambda\left|
      \begin{array}{c}
         (1-n,\; 1)             \\
         (k,\; 1/\mu)
      \end{array}
   \right]\right.
   \quad\mbox{for }0<\mu<1.
\end{equation}
A reparametrization of Eqs.~(11) and (12) using $\langle n\rangle$
in place of the scale parameter $\lambda$ is more convenient.
The factorial moments of $P_n$ are equivalent to the ordinary
moments of $\psi(z)$ for Eq.~(1) therefore
$\langle n\rangle=\langle z\rangle$ and Eq.~(3) yields
\begin{equation}
        \lambda=\frac{\Gamma(k+1/\mu)}
	{\Gamma(k)\,\langle n\rangle}.
\end{equation}
For unit scaling exponent $\mu$ the HNBD reduces to the pure NBD given by
Eq.~(4) which is the $\mu=1$ marginal case of Eq.~(11) for
$\langle n\rangle<k$ and of Eq.~(12) for $\langle n\rangle>k$. This can be
deduced by comparing Eq.~(10) with (A.9) and recalling that
${\sf{}_1F_0}(a,-\,;x)$ yields the binomial function $(1-x)^{-a}$
for $|x|<1$.

At first glance it may seem strange that a three-parameter
discrete probability law requires three different expressions for $P_n$
each having a restricted domain of validity over
the parameter space. But this complication is a reasonable price
for the high degree of flexibility of the HNBD.
It involves as a special or limiting
case the Poisson transform of many widely used probability densities
such as the
\begin{itemize}
\item
gamma distribution for $\mu=1$ and $k>0$,
\item
Weibull distribution for $k=1$ and $\mu>0$,
\item
chi distribution for $\mu=2$ and $k>0$,
\item
Pareto distribution for $k\to0$ and $\mu\to\infty$,
\item
log-normal distribution for $\mu\to0$ and $k\to\infty$
\end{itemize}
to mention but a few.
Therefore one may think the HNBD as a family of different discrete
probability laws rather than a single distribution.
Since the infinite divisibility of $\psi(z)$ is preserved by
$P_n$ for Eq.~(1) we can use Bondesson's theorem~[11] to
deduce that the HNBD is infinitely divisible if $0\leq\mu\leq1$,
otherwise this feature of $P_n$ is violated.

Before turning our attention to data analysis
let us consider briefly the computational aspects of the HNBD.
Unfortunately, the Fox function is not
available in software packages for the computation
of special functions. A straightforward
way to calculate $P_n$ in fitting procedures is the evaluation
of the integral Eq.~(1) numerically with $f(x)$ given by Eq.~(2),
requireing $\langle z\rangle=1$ and changing $z$ to
$\langle n\rangle z$ in the Poisson weight.
For $\mu\to0$ and $k\to\infty$ the HNBD converges to the
Poisson transformed log-normal distribution
which lacks a representation in terms of known functions.
To include this important limiting case into the fits
Eq.~(2) should be written in the following form~[7]:
\begin{equation}
        \psi(z)=\frac{|p|}{\Gamma(p^{-2})\,\sigma z}\,
        \exp\left[p^{-2}w-e^w\right]\qquad\mbox{for $p\neq0$}
\end{equation}
where $w=p\,(\ln z+\alpha)/\sigma$
with new parameters $\sigma>0$, $p$ and $\alpha$.
This reparametrization of the generalized gamma density
allows $\mu<0$ as well ($\mu=p/\sigma$) and
the log-normal distribution can be mapped to the origin in~$p$,
\begin{equation}
        \psi(z)=\frac{1}{\sqrt{2\pi}\,\sigma z}\,
        \exp\left[-\frac{(\ln z-\alpha)^2}
        {2\sigma^2}\right]\quad\mbox{\ \ for $p=0$}.
\end{equation}
Restricting the location parameter $\alpha$ to
\begin{equation}
        \alpha=\left\{\begin{array}{ll}
  \ln\Gamma(p^{-2}+\sigma/p)-\ln\Gamma(p^{-2}) &\qquad\mbox{for $p\neq0$} \\
  -\sigma^2/2                                   &\qquad\mbox{for $p=0$}
        \end{array}\right.
\end{equation}
$P_n$ is computed by inserting Eqs.~(14) and (15)
into Eq.~(1), changing $z$ to $\langle n\rangle z$ in the Poisson weight
and evaluating the integral by the Romberg method or Gaussian
quadrature~[12] or using the numerical integration
package of Mathematica~[13].

Applying the HNBD to experimental data we shall consider typical
reactions and energies where the pure NBD is claimed to be in contradiction
with observations. Significant deviations from $\mu=1$ are expected in
\begin{enumerate}
\item
inelastic $pp$ reactions up to ISR energies,
\item
deep-inelastic $e^+p$ scattering at HERA energies,
\item
$e^+e^-$ annihilations at LEP energies,
\item
non-diffractive $p\bar p$ collisions at top SPS energy $\sqrt s=900$ GeV.
\end{enumerate}
\nopagebreak
Items 2 and 3 have already been investigated in refs.~[7,8] and here only
a few additions will be made. But let us first consider inelastic $pp$
reactions, perhaps the earliest typical example for the failure of
NBD fits. Without the aim of completeness
we have analysed multiplicity distributions in the c.m. energy
range $\sqrt s=10.7-546$~GeV~[14] including the
inelastic $p\bar p$ data of the UA5 Collaboration.
In the fitting procedure numerical evaluation of the integral Eq.~(1)
was carried out as described above. After some initial playing
with the parameters it was found that the
$k=1$ special case of the HNBD,
i.e. the Poisson transformed Weibull distribution, provides good
description of the investigated data. The results of fits are
collected in Table~1. It is worth noticing that the best-fit value of
$\mu$ slowly decreases
with increasing energy, further, $\mu>1$ for each data set.
According to some preliminary results the $k=1$ special case
produces similar quality fits
for inelastic  $\pi^\pm p$ and $K^\pm p$ reactions.

The Poisson transformed Weibull law works
successfully in deep-inelastic $e^+p$ scattering at HERA energies
as well. In ref.~[8] the multiplicity distributions measured by
the H1 Collaboration~[15] were compared to the
Weibull case of the HNBD neglecting systematic errors of $P_n$.
To avoid possible misunderstanding
we repeated the analysis with the inclusion of both sources of
experimental uncertainties. The average multiplicity
$\langle n\rangle$ was fixed at its observed value and only $\mu$ was
treated as free parameter in the fitting procedure.
The results are collected in Table~2,
the quoted errors of $\mu$ are statistical and systematic. The success of
the $k=1$ special case of the HNBD (even with omitted systematic
errors of $P_n$) suggests that the inelastic $pp$ and deep-inelastic
$e^+p$ multiplicity distributions are similarly shaped. This seems to be
confirmed by the observation~[15] that the H1 data and the
best-fit NBD deviate from each other at small multiplicities.
Due to the diffractive fraction of events, the same happens
for inelastic $pp$ data too. In both reactions
the rise of $P_n$ at small~$n$ is less rapid than that of the
NBD and this property is well described by the
Poisson transform of the Weibull distribution with $\mu>1$.

A completely different type of deviation from $\mu=1$ arises in
non-diffractive $p\bar p$ collisions at $\sqrt s=900$~GeV.
The multiplicity distributions measured by the UA5 Collaboration~[16]
are at present the highest energy published data for $P_n$.
The Negative Binomial fits are satisfactory in narrow pseudorapidity
intervals but for $|\eta|>2.5$ the NBD fails to reproduce the observed
shape of $P_n$. According to widespread opinion the failure of the NBD
is caused by the much quoted shoulder structure. In narrow
bins this is not visible but for $|\eta|>2.5$ some kind of structure can
indeed be seen in the tail of $P_n$. However, the dominant source of
discrepancy between the NBD and observations has nothing to do with the
shoulder. It was demonstrated already in~[16] that the real difficulty
for wide $\eta$-intervals is the pronounced narrow peak of the
heavy tailed distributions. Performing HNBD fits
reveals clearly that the $\mu=1$ special case is unable to reproduce a
highly skewed shape which is so characteristic of the UA5 data
for $|\eta|>2.5$. But letting $\mu$ to vary freely in the fits it turns out
that the $\mu\to0$, $k\to\infty$ limiting case of the HNBD, i.e. the Poisson
transformed log-normal distribution, yields reasonable description of $P_n$
possessing a narrow peak and extended tail. The results for
full phase-space and wide $\eta$-windows
are collected in Table~3.
Although systematic deviations between
the experimental and theoretical $P_n$ inevitably occur due to the
shoulder structure (the HNBD is unimodal) the overall shape of the
multiplicity distributions can be reproduced remarkably well by the
two-parameter Poisson transformed log-normal law. For the
full phase-space data the fit is illustrated in Fig.~1.

Interestingly, the log-normal limit of the HNBD was found to be successful
also in $e^+e^-$ annihilations at the $Z^0$ peak. In ref.~[7] the results of
three-parameter HNBD fits were presented favouring $\mu\approx0$
for the full phase-space multiplicity distributions.
Here we provide the outcome of two-parameter Poisson
transformed log-normal fits, see Table~4. The analysed data
sets now include the Aleph data with updated systematic
errors and the Opal data corresponding to $\sqrt s=161$~GeV~[17].
As is seen the quality of fits are satisfactory for each
experiment. Similar results can be obtained for the Aleph data measured
in central rapidity windows~[18], these are quoted in Table~5. We have to
mention that the Delphi data in $y$-intervals exhibit significant
shoulder structure which can not be reproduced by unimodal distributions
and therefore the HNBD fails as well.

The importance of the log-normal distribution to approximate $P_n$ in
$e^+e^-$ and $p\bar p$ collisions has already been stressed in the
literature~[19]. One may think that the observed success of the Poisson
transformed log-normal law is only another manifestation of the above
cited results. But the widely known and accepted evidences for the
log-normality of $P_n$ at present energies are not rarely questionable.
Their comparison to our findings (which indicate that log-normality
emerges asymptotically) will be treated in a separate paper~[20].

In conclusion, we have investigated the H-function extension of the NBD
obtained by the Poisson transform of the generalized gamma
distribution~Eq.~(2). The analytic form of $P_n$ was rederived
via Mellin transform. Since Eq.~(2) involves as special
and limiting cases many classical probability densities, the HNBD is
expected to obey high degree of flexibility. Our applications confirmed
this expectation. Fitting the HNBD to various multiplicity data
claimed to be in disagreement with the pure NBD we have obtained
satisfactory results. According to these findings, two major types of
departure from $\mu=1$ arise in multiparticle production. For inelastic
$pp$ and deep-inelastic $e^+p$ scattering the dominant source of
discrepancy is the less steep rise of $P_n$ at small multiplicities
caused by the diffractive component of the underlying dynamics
(at least for $pp$ data).
The multiplicity distributions can be reproduced successfully by
the Weibull case of the HNBD with $\mu>1$.
The other $\mu\neq1$ type behaviour arises in non-diffractive
$p\bar p$ collisions at $\sqrt s=900$~GeV where the NBD is incompatible with
the highly skewed shape of $P_n$ for $|\eta|>2.5$. These heavy tailed
distributions with a narrow peak are best described if
$\mu\to0$ and $k\to\infty$, i.e. by the log-normal limit of the HNBD.
Despite of the absence of skew multiplicity curves,
the same type of departure from the pure NBD has been
observed also in $e^+e^-$ annihilations at LEP energies.
This remarkable ubiquity of the $\mu\to0$ limit
for the $e^+e^-$ and $p\bar p$ data at top energies
has a natural explanation based on renormalization group arguments
for asymptotic KNO scaling~[20]. It will be interesting to see how
precisely can the Poisson transformed log-normal law reproduce the
forthcoming $p\bar p$ data at $\sqrt s=1800$ GeV at Tevatron.

The observed success of the HNBD indicates
that the majority of existing multiplicity data
can be interpreted on the basis of the asymptotic
KNO scaling form of the Negative Binomial Distribution
with the extension of its validity to positive powers of the
scaling variable. It is hoped that in the next 25 years
the HNBD will provide a helpful unifying framework to study the NB
regularity and the origin of possible deviations.

\section*{Acknowledgements}

I am indebted to T. Cs\"org\H o for the helpful comments on the
manuscript and to G. Jancs\'o for the constructive criticisms.
This work was supported by the Hungarian Science Foundation under
Grant No. OTKA-T024094/1997.

\newpage
\begin{center}
{\bf Appendix: \ The H-function of Fox}
\end{center}
The Fox function ${\sf H^{m,n}_{p,\;q}}(x)$ is defined by the
Mellin-Barnes type integral
\begin{equation}
   {\sf H^{m,n}_{p,\;q}}\left[\,x\left|
      \begin{array}{c}
         (a_p, \alpha_p)      \\
         (b_q, \beta_q)
      \end{array}
   \right]\right.=
   {\sf H^{m,n}_{p,\;q}}
   \left[x\left|
      \begin{array}{c}
         (a_1, \alpha_1),\ldots,(a_p, \alpha_p)       \\
         (b_1, \beta_1),\ldots,(b_q, \beta_q)
      \end{array}
   \right]\right.=
   \frac{1}{2\pi i}\int_{\cal C}\,\frac{A(s)B(s)}{C(s)D(s)}\,x^s\,ds
   \eqnum{A.1}
\end{equation}
with $x\neq0$ and
\begin{equation}
\begin{array}{ll}
	A(s)=\prod_{j=1}^m\Gamma(b_j-\beta_js),\quad		&
	B(s)=\prod_{j=1}^n\Gamma(1-a_j+\alpha_js),		\\
								&
								\\
	C(s)=\prod_{j=m+1}^q\Gamma(1-b_j+\beta_js),\quad	&
	D(s)=\prod_{j=n+1}^p\Gamma(a_j-\alpha_js).\eqnum{A.2}
\end{array}
\end{equation}
Above $m,n,p,q$ are integers satisfying $0\leq n\leq p$ and $1\leq m\leq q$.
In the cases $n=0$, $q=m$ and $p=n$ (A.2) has to be interpreted as
$B(s)=1$, $C(s)=1$ and $D(s)=1$, respectively.
The parameters $(a_1,\ldots,a_p)$ and $(b_1,\ldots,b_q)$ are complex,
whereas parameters $(\alpha_1,\ldots,\alpha_p)$
and $(\beta_1,\ldots,\beta_q)$ are
positive numbers. They are restricted by the condition that
        $\alpha_j(b_h+\nu)\neq\beta_h(a_j-1-\lambda)$
for $\nu,\lambda=0,1,\ldots$; $h=1,\ldots,m$; $j=1,\ldots,n$. The
contour ${\cal C}$ in the complex $s$ plane is such that the points
$s=(b_h+\nu)/\beta_h$ and $s=(a_j-1-\nu)/\alpha_j$ lie to the
right and left of ${\cal C}$ respectively while ${\cal C}$ extends from
$s=\infty-ik$ to $s=\infty+ik$ where $k$ is a constant with
$k>|{\rm Im\ } b_h|/\beta_h$.

The Fox function makes sense
only if the following two existence conditions are satisfied:
\begin{enumerate}
\item
$x\neq0$ and $\kappa>0$ with
$\kappa=\sum_{j=1}^q\beta_j-\sum_{j=1}^p\alpha_j$
\item
$\kappa=0$ and $0<|x|<1/\rho$ with
$\rho=\prod_{j=1}^p\alpha_j^{\alpha_j}\prod_{j=1}^q\beta_j^{-\beta_j}$.
\end{enumerate}
Under these conditions ${\sf H}(x)$ is an analytic function for
$x\neq0$, in general multivalued, one-valued on the Riemann
surface of $\ln x$.

Elementary properties of the Fox function very useful
to manipulate them are
\begin{equation}
        {\sf H^{m,n}_{p,\;q}}\left[\,\frac{1}{x}\left|
        \begin{array}{c}
                 (a_p, \alpha_p)      \\
                 (b_q, \beta_q)
        \end{array}
        \right]\right.=
        {\sf H^{n,m}_{q,\;p}}\left[\,x\left|
        \begin{array}{c}
                 (1-b_q,\; \beta_q)     \\
                 (1-a_p,\; \alpha_p)
        \end{array}
        \right]\right.
        \eqnum{A.3}
\end{equation}
\begin{equation}
        x^c\,{\sf H^{m,n}_{p,\;q}}\left[\,x\left|
        \begin{array}{c}
                 (a_p, \alpha_p)      \\
                 (b_q, \beta_q)
        \end{array}
        \right]\right.=
        {\sf H^{m,n}_{p,\;q}}\left[\,x\left|
        \begin{array}{c}
                 (a_p+c\,\alpha_p,\; \alpha_p)          \\
                 (b_q+c\,\beta_q,\; \beta_q)
        \end{array}
        \right]\right.
        \eqnum{A.4}
\end{equation}
\begin{equation}
        {\sf H^{m,n}_{p,\;q}}\left[\,x^c\left|
        \begin{array}{c}
                 (a_p, \alpha_p)      \\
                 (b_q, \beta_q)
        \end{array}
        \right]\right.=
        \frac{1}{c}\,{\sf H^{m,n}_{p,\;q}}\left[\,x\left|
        \begin{array}{c}
                 (a_p,\; \alpha_p/c)            \\
                 (b_q,\; \beta_q/c)
        \end{array}
        \right]\right.\quad c>0.
        \eqnum{A.5}
\end{equation}
Many of the integral transforms of ${\sf H}(x)$ yield
Fox functions again with altered parameters. For example,
the Laplace transform is given by the equation pair
\begin{equation}
	{\cal L}\left\{{\sf H^{m,n}_{p,\;q}}(cx),r\right\}=
	\frac{1}{c}\,
        {\sf H^{n+1,m}_{q,\;p+1}}
        \left[\,\frac{r}{c}\left|
        \begin{array}{c}
                 (1-b_q-\beta_q,\; \beta_q)     \\
                 (0,1),\;(1-a_p-\alpha_p,\; \alpha_p)
        \end{array}
        \right]\right.\quad\mbox{for $0\leq\kappa\leq1$}
        \eqnum{A.6}
\end{equation}
and
\begin{equation}
	{\cal L}\left\{{\sf H^{m,n}_{p,\;q}}(cx),r\right\}=
	\frac{1}{c}\,
        {\sf H^{m,n+1}_{p+1,\;q}}
        \left[\,\frac{c}{r}\left|
        \begin{array}{c}
                 (1,1),\;(a_p+\alpha_p,\; \alpha_p) \\
                 (b_q+\beta_q,\; \beta_q)     \\
        \end{array}
        \right]\right.\quad\mbox{for $\kappa\geq1$}
        \eqnum{A.7}
\end{equation}
with scale parameter $c>0$. The Mellin transform of ${\sf H}(x)$ reads
\begin{equation}
	{\cal M}\left\{{\sf H^{m,n}_{p,\;q}}(cx),s\right\}=
	\frac{\prod_{j=1}^m\Gamma(b_j+\beta_js)\,
	      \prod_{j=1}^n\Gamma(1-a_j-\alpha_js)}
	     {\prod_{j=m+1}^q\Gamma(1-b_j-\beta_js)\,
	      \prod_{j=n+1}^p\Gamma(a_j+\alpha_js)}\;c^{-s}.\eqnum{A.8}
\end{equation}
An important special case of ${\sf H}(x)$ is the generalized
hypergeometric function ${\sf{}_pF_q}(x)$. Its relation to the
Fox function is given by
\begin{eqnarray}
      \frac{\prod_{j=1}^p\Gamma(a_j)}{\prod_{j=1}^q\Gamma(b_j)}\;
      &&{\sf{}_pF_q}
      \left[\left.
      \begin{array}{c}
         a_1,\ldots,a_p         \\
         b_1,\ldots,b_q
      \end{array}\right|x\,
      \right]=
      {\sf H^{1,p}_{p,q+1}}
      \left[-x\left|
      \begin{array}{c}
         (1-a_p,1)        \\
         (0,1),\;(1-b_q,1)
      \end{array}
	\right]\right.\nonumber\\=\;&&
	{\sf H^{p,1}_{q+1,p}}
      \left[-\frac{1}{x}\left|
      \begin{array}{c}
         (1,1),\;(b_q,1)	\\
	 (a_p,1)
      \end{array}
\right]\right.
\quad\mbox{for $p\leq q$, or $p=q+1$ and $|x|<1$.}
\eqnum{A.9}
\end{eqnarray}
For more details on the H-function as well as for a rich collection of
its particular cases the reader is referred to [9].

\newpage

\mediumtext
\begin{table}
\begin{tabular}{ccrr}
$\sqrt s$ (GeV) & $\mu$  & $\langle n\rangle\ \ \ \ \ \ \ $
& $\chi^2$/d.o.f. \\
\tableline
 10.7 & $5.353\pm1.431$ & $  5.503\pm0.171$ & $ 1.7/6\ \,$  \\
 13.8 & $3.064\pm0.298$ & $  6.465\pm0.151$ & $ 4.5/8\ \,$  \\
 13.9 & $3.248\pm0.250$ & $  6.258\pm0.112$ & $ 4.5/8\ \,$  \\
 16.6 & $2.765\pm0.087$ & $  6.955\pm0.071$ & $16.3/11$ \\
 18.2 & $2.892\pm0.551$ & $  7.407\pm0.410$ & $ 2.1/8\ \,$  \\
 19.7 & $2.942\pm0.160$ & $  7.603\pm0.128$ & $ 7.8/11$ \\
 21.7 & $2.575\pm0.125$ & $  7.838\pm0.129$ & $15.4/12$ \\
 23.9 & $2.678\pm0.225$ & $  8.735\pm0.224$ & $13.5/11$ \\
 26.0 & $2.725\pm0.113$ & $  9.077\pm0.114$ & $ 8.5/11$ \\
 27.6 & $2.485\pm0.101$ & $  8.869\pm0.136$ & $16.7/14$ \\
 30.4 & $2.468\pm0.150$ & $  9.341\pm0.261$ & $ 5.6/15$ \\
 38.8 & $2.503\pm0.076$ & $ 10.045\pm0.112$ & $11.9/14$ \\
 44.0 & $2.349\pm0.135$ & $ 10.703\pm0.223$ & $ 4.5/17$ \\
 52.6 & $2.329\pm0.109$ & $ 11.457\pm0.212$ & $10.4/19$ \\
 62.2 & $2.392\pm0.105$ & $ 12.267\pm0.258$ & $17.5/18$ \\
$\!\!546.0$ & $1.840\pm0.034$ & $27.255\pm0.292$ & $50.9/45$ \\
\end{tabular}
\medskip
\caption{Results of HNBD fits (Weibull case, $k=1$) to the
multiplicity distributions measured in inelastic $pp$ reactions~[14].
The last row corresponds to the inelastic $p\bar p$ data of the
UA5 Collaboration.}
\end{table}

\bigskip

\begin{table}
\begin{tabular}{crcc}
$\eta^*$-interval &    $W$ (GeV)   &     $\mu$        & $\chi^2$/d.o.f. \\
\tableline
$1<\eta^*<2$      & $  80\div115$  & $1.916\pm0.194$  & $1.6/13$ \\
                  & $ 115\div150$  & $1.872\pm0.085$  & $0.6/14$ \\
                  & $ 150\div185$  & $1.891\pm0.212$  & $1.5/14$ \\
                  & $ 185\div220$  & $1.841\pm0.203$  & $0.8/14$ \\ \tableline
$1<\eta^*<3$      & $  80\div115$  & $2.336\pm0.166$  & $2.5/17$ \\
                  & $ 115\div150$  & $2.116\pm0.143$  & $2.4/18$ \\
                  & $ 150\div185$  & $2.041\pm0.143$  & $0.9/20$ \\
                  & $ 185\div220$  & $2.094\pm0.155$  & $0.6/21$ \\ \tableline
$1<\eta^*<4$      & $  80\div115$  & $3.488\pm0.385$  & $0.8/18$ \\
                  & $ 115\div150$  & $3.100\pm0.180$  & $2.2/20$ \\
                  & $ 150\div185$  & $2.920\pm0.226$  & $1.5/22$ \\
                  & $ 185\div220$  & $2.733\pm0.192$  & $1.6/22$ \\ \tableline
$1<\eta^*<5$      & $  80\div115$  & $4.811\pm0.392$  & $0.9/18$ \\
                  & $ 115\div150$  & $4.238\pm0.277$  & $2.1/21$ \\
                  & $ 150\div185$  & $4.011\pm0.288$  & $1.9/22$ \\
                  & $ 185\div220$  & $4.021\pm0.334$  & $1.5/23$ \\
\end{tabular}
\medskip
\caption{Results of HNBD fits (Weibull case, $k=1$) to the deep-inelastic
$e^+p$ multiplicity data of the H1 Collaboration~[15]. The average
multiplicity $\langle n\rangle$ was fixed at its experimental value.}
\end{table}

\newpage

\begin{table}
\begin{tabular}{lccc}
$\eta$-interval & $\sigma$ & $\langle n\rangle$ & $\chi^2$/d.o.f.\\ \tableline
$\Delta\eta=3.0$  & $0.702\pm0.020$ & $22.050\pm0.546$ & $30.2/81$ \\
$\Delta\eta=5.0$  & $0.621\pm0.018$ & $32.747\pm0.780$ & $35.9/99$ \\
full phase-space& $0.538\pm0.014$ & $35.552\pm0.834$ & $32.7/52$ \\
\end{tabular}
\medskip
\caption{Results of HNBD fits (log-normal case, $p=0$) to the UA5
non-diffractive $p\bar p$ multiplicity data for $|\eta|>2.5$
at $\sqrt s=900$~GeV~[16].
The $(p,\sigma)$ parametrization of the HNBD is discussed in the text.}
\end{table}

%bigskip

\begin{table}
\begin{tabular}{lccr}
Experiment    & $\sigma$ & $\langle n\rangle$ & $\chi^2$/d.o.f. \\ \tableline
Aleph         & $0.202\pm0.008$ & $21.094\pm0.226$ & $11.2/25$  \\
Delphi        & $0.201\pm0.004$ & $21.353\pm0.104$ & $32.4/24$  \\
L3            & $0.210\pm0.009$ & $20.742\pm0.254$ & $15.9/21$  \\
Opal          & $0.213\pm0.005$ & $21.349\pm0.103$ & $17.2/25$  \\
Opal, 161 GeV & $0.235\pm0.016$ & $24.417\pm0.423$ & $ 3.9/23$  \\
\end{tabular}
\medskip
\caption{Results of HNBD fits (log-normal case, $p=0$) to the
$e^+e^-$ full phase-space multiplicity distributions at
$\sqrt s=91$~GeV (top four rows) and 161 GeV~[17].}
\end{table}

%\bigskip

\begin{table}
\begin{tabular}{lcrr}
$y$-interval & $\sigma$ & $\langle n\rangle\ \ \ \ \ \ \ $
& $\chi^2$/d.o.f.\\ \tableline
$\Delta y=0.5$  & $0.565\pm0.046$ & $ 3.058\pm0.143$ & $ 3.5/18$  \\
$\Delta y=1.0$  & $0.547\pm0.030$ & $ 6.426\pm0.231$ & $ 9.3/32$  \\
$\Delta y=1.5$  & $0.505\pm0.018$ & $ 9.897\pm0.219$ & $26.3/38$  \\
$\Delta y=2.0$  & $0.454\pm0.021$ & $13.284\pm0.293$ & $13.7/44$  \\
\end{tabular}
\medskip
\caption{Results of HNBD fits (log-normal case, $p=0$) to the
Aleph multiplicity data in central rapidity windows~[18].}
\end{table}

\bigskip

\begin{figure}
\hbox{\hskip0cm
\epsfig{file=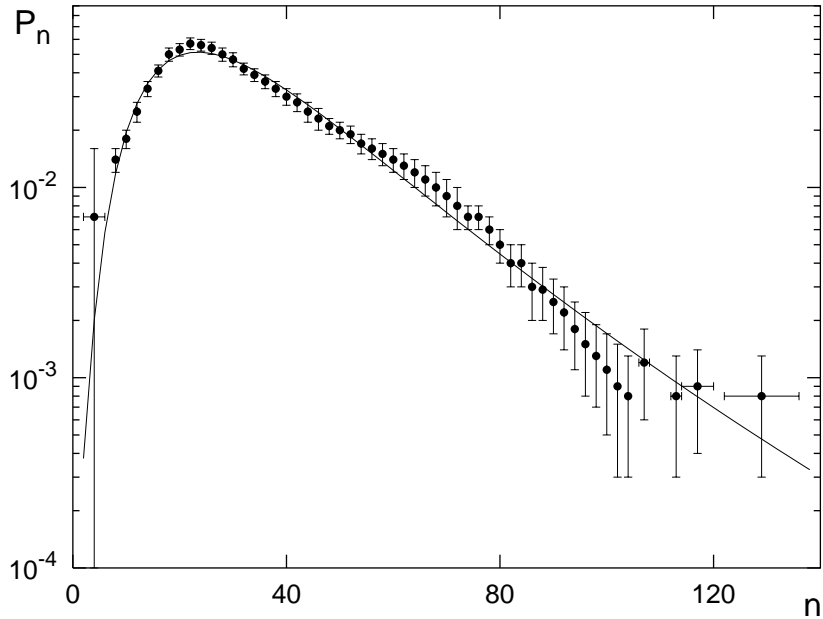}}
\bigskip
\caption{HNBD fit (lognormal case, $p=0$) to the non-diffractive $p\bar p$
multiplicity distribution in full phase-space at $\sqrt s=900$ GeV.}
\end{figure}

\end{document}